\documentclass[twocolumn,pra,showpacs,floatfix,amsmath,amsfonts]{revtex4}
\usepackage[latin9]{inputenc}
\usepackage{color}
\usepackage{amsmath}
\usepackage{graphicx}
\usepackage{esint}

\usepackage[colorlinks]{hyperref}

\makeatletter
\@ifundefined{textcolor}{}
{%
 \definecolor{BLACK}{gray}{0}
 \definecolor{WHITE}{gray}{1}
 \definecolor{RED}{rgb}{1,0,0}
 \definecolor{GREEN}{rgb}{0,1,0}
 \definecolor{BLUE}{rgb}{0,0,1}
 \definecolor{CYAN}{cmyk}{1,0,0,0}
 \definecolor{MAGENTA}{cmyk}{0,1,0,0}
 \definecolor{YELLOW}{cmyk}{0,0,1,0}
}

\usepackage{color}
\usepackage{graphics}
\usepackage{epsfig}
\newcommand{\be}{\begin{equation}}
\newcommand{\ee}{\end{equation}}
\newcommand{\bea}{\begin{eqnarray}}
\newcommand{\eea}{\end{eqnarray}}
\newcommand{\bes}{\begin{subequations}}
\newcommand{\ees}{\end{subequations}}

\newcommand{\tx}{\tilde{x}}

\newcommand{\tpsi}{\tilde{\psi}}
\newcommand{\tmu}{\tilde{\mu}}

\newcommand{\bV}{\mathbf{V}}

\newcommand{\bpsi}{\mbox{\boldmath$\psi$\unboldmath}}

\newcommand{\bPsi}{\mathbf{\Psi}}

\newcommand{\bic}{{\rm bic}}

\hypersetup{backref,colorlinks=true,linkcolor=blue,citecolor=blue,urlcolor=blue,linktocpage=true,breaklinks}

\makeatother

\begin{document}

\title{Bound States in the Continuum in Spin-Orbit Coupled Atomic Systems}

\author{Yaroslav V. Kartashov$^{1,2}$, Vladimir V. Konotop$^{3}$,  and Lluis Torner$^{1,4}$ }

\affiliation{$^{1}$ ICFO-Institut de Ciencies Fotoniques, The Barcelona Institute
of Science and Technology, 08860 Castelldefels (Barcelona), Spain
\\
 $^{2}$Institute of Spectroscopy, Russian Academy of Sciences, Troitsk,
Moscow Region, 142190, Russia \\
 $^{3}$ Centro de F\'{i}sica Te\'orica e Computacional and Departamento de F\'{i}sica, Faculdade de Ci\^encias, Universidade
de Lisboa, Campo Grande, Ed. C8, Lisboa 1749-016, Portugal}

\affiliation{$^{4}$Universitat Politecnica de Catalunya, 08034 Barcelona, Spain }

\date{\today}
\begin{abstract}

We show that the interplay between spin-orbit coupling and Zeeman splitting in atomic systems can lead to the existence of bound states in the continuum (BICs) supported by trapping potentials. Such states have energies falling well within the continuum spectrum, but nevertheless they are localized and fully radiationless. We report the existence of BICs, in some cases in exact analytical form, in systems with tunable spin-orbit coupling and show that the phenomenon is physically robust. We also found that BIC states may be excited in spin-orbit-coupled Bose-Einstein condensates, where under suitable conditions they may be metastable with remarkably long lifetimes.

\end{abstract}


\maketitle

Systems of cold trapped atoms are of paramount current importance for both, fundamental physics and future quantum technologies. Salient examples include quantum simulators~\cite{simulators, maciej}, or the realization of synthetic magnetic~\cite{magnetic} and electric~\cite{electric} fields as well as nearly arbitrary gauge potentials~\cite{multi-level}. Multilevel atomic systems allow for the exploration of spin-orbit coupling (SOC)~\cite{Dressel,Rashba} at the macroscopic scale. SOC is a fundamental mechanism that couples the momentum and spin degrees of freedom, thereby introducing a wealth of physical phenomena. For example, SOC~\cite{SO-BEC} may change the spectrum of the system to an extent that stripe-phase states emerge~\cite{stripe}, it modifies the band structure of periodic potentials thus strongly affecting Bloch oscillations~\cite{Bloch}, and introduces new  properties in nonlinear states~\cite{solitons1,solitons2} supported by Bose-Einstein condensates (SO-BECs). However, to date  attention has only been devoted to conventional localized modes. Here we uncover the existence of BICs, i.e., discrete localized states with energies falling into the continuous spectrum that contrary to intuition are radiationless, in atomic systems with SOC.

The mathematical existence of BICs is known since the early days of quantum mechanics, as the first example was constructed by von Neumann and Wigner in 1929~\cite{first}. Several decades later, the concept was revisited~\cite{Stillinger} and extended to coupled-resonances~\cite{Friedrich}, and subsequently several approaches to construct potentials supporting BICs were reported, including schemes based on Darboux transformations~\cite{Svirskyt}, super-symmetric quantum mechanics~\cite{SUSY}, and Hamiltonian separability~\cite{Robnik86}, among others (see~\cite{BIC_review}).  Signatures of BICs were observed experimentally already in the 60s in acoustics~\cite{parker}, but the topic has gained important momentum after BICs were observed in recent experiments in optical systems~\cite{opt1}. During the last years it has been shown that different mechanisms may lead to the formation of BICs~\cite{BIC_review}. For example, they can arise in lattices with engineered hopping strengths, in systems where coupling of continuous modes of different symmetry is prohibited, via collapse of Fano resonances, as embedded eigenvalues in nanostructures, as surface states, in waveguides with defects,  at the interface of topologically different materials, or in fully-vectorial anisotropic media\cite{Lad10,2D,opt2,opt3,topology,anisotropy}. Practical applications have been suggested in acoustics~\cite{waterwaves}; spintronics~\cite{spintronics}, for spin filters~\cite{Vall10} and spin-polarized devices~\cite{RamOre14}; and photonics, for integrated opto-electronic devices~\cite{Capasso-integrated}, ultrahigh-Q resonances in subwavelength films~\cite{ultrahighQ}, and super-cavity lasing~\cite{BIC-laser}.

In this paper we uncover the existence of BICs in a whole new class of physical systems, constituted by two-level atoms, including BECs, with SOC. We find analytical expressions for BIC states in selected trapping potentials and, importantly, show that the phenomenon is robust, i.e., that BICs also exist in potentials deviating from the analytical ones. Also, calculations for nonlinear systems with two-body interactions suggest the existence of almost radiationless BICs in BECs with SOC, too.

We start by addressing a two-level atom placed in an external (synthetic) magnetic field inducing a Zeeman splitting $2\Omega$. The atomic levels 1 and 2 are affected by the potentials $V_1(x)$ and $V_2(x)$, respectively. We address a one-dimensional geometry and account for SOC in the form $\gamma k \sigma_y$~\cite{Dressel,Rashba}, with $\gamma$ being the SOC strength. In the  units where  $\hbar=m=1$, the Hamiltonian reads
\begin{equation}
\label{Ham}
 H=  \frac{k^{2}}{2}+ \gamma k\sigma_y  +  \Omega\sigma_{z} +  \bV, \,\,\, \bV=\left(\begin{array}{cc} V_1(x) &0 \\ 0 & V_2(x)
 \end{array}\right),
\end{equation}
where $k=-i\partial_x$    and  $\sigma_{x,y,z}$ are the Pauli matrices. It is assumed that the potentials $V_{1,2}(x)$ have no singularities and that they decay at $x\to\pm \infty$ exponentially or faster, i.e., $\lim_{x\to\pm\infty}V_{1,2}(x)=0$.

First we present qualitative arguments showing that a broad class of potentials $\bV(x)$, satisfying quite general conditions, may support BICs in the presence of SOC. To this end, we recall that in the absence of the potential (i.e., when $\bV(x)\equiv 0$) the dispersion relation, defined by
 $H \bpsi =\mu(k)\bpsi$,
where $\bpsi\equiv (\psi_1 (k,x), \psi_2(k,x))^T, $
consists of two branches: $\mu_\pm(k) = k^{2}/2\pm\sqrt{\Omega^2+\gamma^{2} k^2}$. Consider now weak potentials acting on each component,{\color{black} which allow for the scaling:} $V_{1,2}(x)=\epsilon^2U_{1,2}(\tx)$, where $U_{1,2}(\tx)=\mathcal{O}(1)$, $\epsilon\ll \Omega$ is a small parameter, and $\tx=\epsilon x$. Localized states associated with discrete energy levels (if any) bifurcate from the above mentioned branches of the continuous spectrum. We are interested in states bifurcating from the bottom of the upper branch $\min_k\mu_+=\mu_+(0)=\Omega$. We thus look for a solution in the form $\bpsi=( \tpsi_1(\tx),\epsilon \tpsi_2(\tx))^T$, where $\tpsi_{1,2}\sim \mathcal{O}(1)$ and the energy can be approximated as $\mu\approx \Omega+\epsilon^2\tmu$. Then, the equation for the second component yields the relation $\tpsi_2 \approx \gamma  \partial_{\tilde{x}}\tpsi_1$ and one obtains for the first component: $-[(\Omega^2+\gamma^2)/2]\partial_{\tx}^2\tpsi_1+U_1(\tx)\tpsi_1=\tmu\tpsi_1 $. This is the Schr\"odinger equation for a particle with the effective mass $1/(\Omega^2+\gamma^2)$. Thus, if the potential $U(\tx)$ has at least one discrete level at negative energy, $\tmu<0$, this level will be embedded in the continuous spectrum associated with the lower branch of the dispersion relation, since $\Omega-\epsilon^2\tmu>\min_k\mu_-(k)$, and will correspond to a BIC. This analysis leads to two important conclusions. First, {\em Zeeman splitting, opening a semi-gap, i.e. a frequency range where only one branch of the spectrum exists, is of crucial importance for the BICs} considered below; therefore we will consider only nonzero splitting. Second, a BIC, if it exists, appears in the gap between the minima of two branches of the continuous spectrum, i.e.,  $\min_k\mu_-(k)<\mu_\bic<\Omega$.

The above observations are valid also for finite-amplitude  potentials provided that they go to zero at infinity fast enough, as shown below both, analytically and numerically. Assuming that $\bV(x)$ decays faster than exponentially at $|x|\to\infty$, while the decay of a BIC (if any) is exponential, i.e., $\bpsi_\bic\sim e^{-\nu |x|}$ at $x\to\infty$, one can establish an explicit relation between the decay exponent $\nu$, the BIC energy, $\mu_\bic$, and the momentum $k_\bic$ of the extended state having the same energy
as the BIC. In our case there exist two bound states with the same exponent $\nu$ and different energies $\tmu_\pm(\nu)=-\nu^2/2\pm\sqrt{\Omega^2-\nu^2\gamma^2}$.  A BIC may exist in a system in which for a decay exponent $\nu$ at least one of $\mu_\pm$ satisfies the condition $\min_k\mu_-(k)<\tmu_\pm(\nu)<\Omega$. Notice that depending on the SOC strength, $\mu_-(k)$ may have either one or two minima: if $0\leq \gamma^2 < \Omega$, there is only minimum $\mu_-(0)=-\Omega$ at $k=0$, while if $\gamma^2>\Omega$ there are two minima  $\mu_-(\pm k_{\rm min})=-(\Omega^2/\gamma^2+\gamma^2)/2 $, where $k_{\rm min}=\left(\gamma^2-\Omega^2/\gamma^2\right)^{1/2}$.  The above condition for the existence of a BIC is always satisfied when
     \begin{eqnarray}
    \label{mu_bic}
    \mu_\bic=\mu_-(k_\bic)=\tmu_+=-\nu^2/2+\sqrt{\Omega^2-\nu^2\gamma^2}
    \end{eqnarray}
which means that a BIC with the corresponding energy exists for any nonzero Zeeman splitting and SOC strength. Moreover, for a strong enough SOC such that $\gamma^2>\Omega/2$, a second BIC with lower energy $\tmu_- >-\Omega$ becomes possible. Such a second BIC delocalizes at the energy of the bottom of the spectrum if $\gamma^2\geq \Omega$. If $\Omega/2<\gamma^2<\Omega$ the lower energy BIC remains localized. Formula (\ref{mu_bic}) reveals the relevance of SOC for a BIC existence: {\em SOC's strength is a tunable parameter ensuring matching conditions for decaying radiationless state.}

\begin{figure}
 	\includegraphics[width=1\columnwidth]{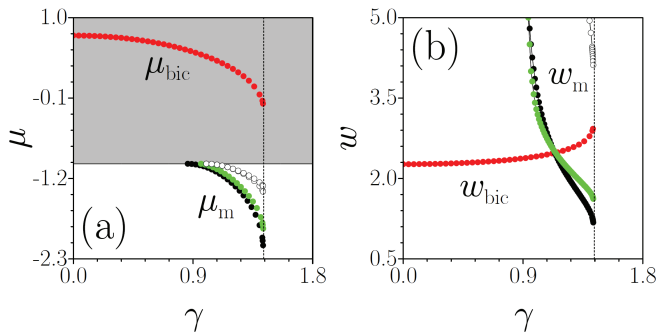}
 	
 	\caption{(a)  BIC energy  (red dots) and  the discrete spectrum $\mu_m$ of (\ref{potV})with $\nu =0.7$ vs SOC strength. The gray region corresponds to the continuous spectrum. (b) Integral widths of the BIC $w_\bic$ and of the states of the discrete spectrum $w_m$ vs SOC strength.  The vertical dashed lines indicate the value $\gamma=1/\nu$, beyond which the potential  (\ref{potV}) ceases to exist. {\color{black} Note that even though at $\gamma=0$ the localized mode of the system can be found, it is not a BIC anymore, because in this limit the two components of the spinor become decoupled and have independent spectra.} }
 	\label{fig:one}
 \end{figure}

\begin{figure}
 	\includegraphics[width=1\columnwidth]{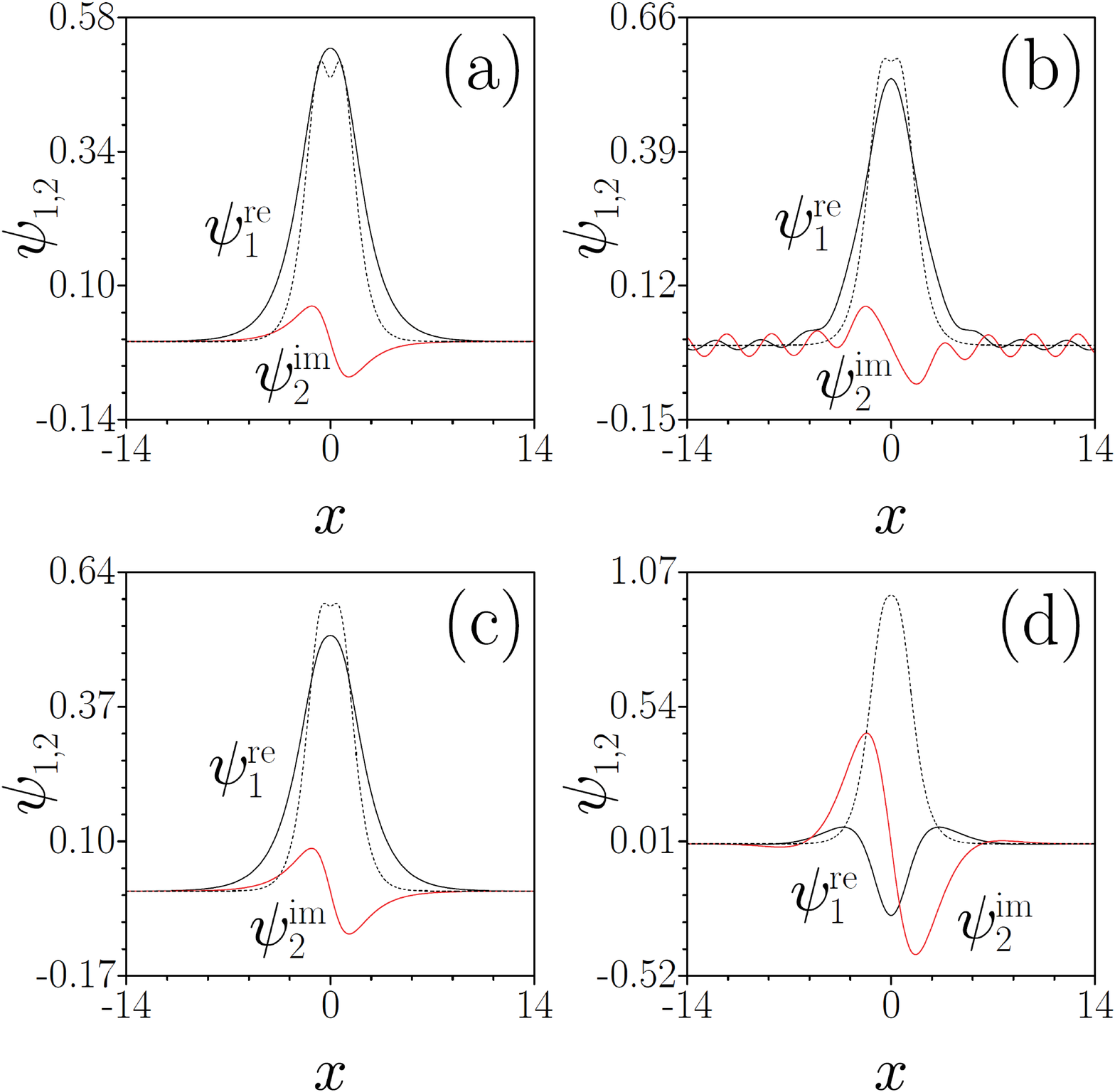}
 	\caption{(a) The potential (\ref{potV}) and the BIC (\ref{BIC}) at $\gamma=0.5$, $\nu=0.7$. (b) State with oscillating tails  in  $V_{\rm pert}$ with  $v_0=0.1$ at $\gamma=0.5$ and $\nu=0.7$.  (c) BIC in a perturbed
 		potential obtained for $\gamma=0.723$. (d) Usual guided mode supported
 		by the designed potential at $\gamma=1$.  The dashed lines in all panels show the inverted potential $-V(x)$. }
 	\label{fig:two}
 \end{figure}

Now we rigorously construct an exponentially-localized potential for which an exact analytical solution can be obtained. We set $V_2(x)\equiv 0$. As suggested by the above perturbative approach, exact solutions can be constructed using the following algorithm: For a chosen shape of the component $\psi_2$  [for which the exponential asymptotics
define the BIC energy (\ref{mu_bic})] we obtain the component $\psi_1$ from the Schr\"odinger equation for $\psi_2$, viewed as an inhomogeneous first-order ordinary-differential equation for $\psi_1$. Using the found spinor $\bpsi$, we derive the exact potential $V_1(x)$ supporting the BIC from the equation for the first component. The algorithm allows for the construction of an infinite number of different potentials supporting BICs and can be used for both types of BICs discussed above. From now on here we study in detail only high-energy BICs with $\mu_\bic$ given by (\ref{mu_bic}). A potential supporting a high-energy BIC is given by (we set $\Omega=1$ without loosing generality):
 	\begin{equation}
 \label{potV}
 V_1=\frac{ 2\nu^2[3\nu^2-(\gamma^2+2\nu^2+\sqrt{1-\gamma^2\nu^2}) \cosh^2(\nu x)]}{\cosh^2(\nu x)[(1+\sqrt{1-\gamma^2\nu^2})\cosh^2(\nu x)-\nu^2]}.
 \end{equation}
This potential has two free parameters: the SOC strength $\gamma$ and the transverse decay rate of the mode $\nu>0$, which ought to satisfy $\nu\gamma<1$. The exact {\color{black} normalized} BIC spinor for such potential reads
 \begin{equation}
 \label{BIC}
 \bpsi_\bic=\frac{C}{\cosh(\nu x)}
\left(\!\begin{array}{c}
 1+\sqrt{1-\gamma^2\nu^2}-\nu^2\cosh^{-2}(\nu x)
 \\
 -\gamma\nu \tanh(\nu x)
\end{array}\!\right)
 \end{equation}
 {\color{black} where the tion constant is defined by requiring $ N=\int_{-\infty}^{\infty} \bpsi^\dag \bpsi dx$ to be one, $N=1$, and for the spinor (\ref{BIC}) is computed as
 	\begin{eqnarray}
 	\frac{1}{C^2}=\frac{4}{3}\left[\left(3-2\nu^2\right)\left(1+\sqrt{1-\gamma^2\nu^2}\right)+\frac{4\nu^4}{5}-\nu^2\gamma^2\right].
 	\end{eqnarray}
 	We also notice that in the limit $\gamma=0$, i.e. when SOC is absent, the system becomes decoupled and in the solution (\ref{BIC}) $\psi_2=0$, while $\psi_1$ represents an ordinary bound state of the potential (\ref{potV}) also computed at $\gamma=0$.
}
The BIC, whose energy as a function of $\gamma$ is shown by the red dots in Fig.~\ref{fig:one}(a), is the only bound state of the potential (\ref{potV}) for small values of $\gamma$, while for large values of $\gamma$ the BIC coexists with localized modes from the standard discrete spectrum (see branches with black, green, and open dots bifurcating from the lower edge of the continuous spectrum). The critical SOC strength $\gamma_{\rm bif}$, at which the bifurcation of the first mode from the discrete spectrum occurs is very close to the value $\gamma=0.8762$, at which the potential changes its shape from a double-well to a single-well [see Figs.~\ref{fig:two} (a) and (d) that depict, respectively, the profiles of the BIC and a mode from the discrete spectrum]. In Fig.~\ref{fig:one} we show the energies of only the simplest modes from the discrete spectrum, but more of them appear when $\gamma$ approaches $1/\nu$. The widths of such modes, defined here as $w=(2/N)\int_{-\infty}^{\infty} \bpsi^\dag |x|\bpsi dx$
rapidly decrease with increasing $\gamma$, while the BIC broadens instead, as shown in Fig.~\ref{fig:one}(b).
 When an exact potential is perturbed, $V_{\rm pert}=V(x)+v(x)$, and everything else remains constant, the BIC ceases to exist,
 as it transforms into a state with small oscillating tails. An  example  corresponding to the potential (\ref{potV}) with an added Gaussian perturbation $v(x)=v_0\exp(-x^2)$, where $v_0\ll 1$, is shown in Fig.~\ref{fig:two}(b).

\begin{figure}[t]
\includegraphics[width=1\columnwidth]{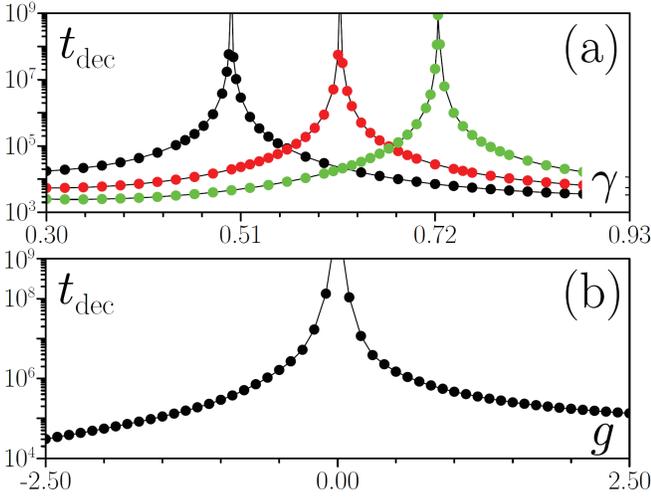}
\caption{ (a) Lifetime of dynamically excited states versus SOC strength for the potential (\ref{potV}) where the BIC exists at $\gamma=0.5$ and $\nu=0.7$ (black circles), and for the perturbed potential $V_{\rm pert}(x)$ with $v_0=0.05$ (red circles) and $v_0=0.1$ (green circles) in the linear case. (b) Lifetime in the potential (\ref{potV}) with $\gamma =0.5$, $\nu =0.7$ vs nonlinearity strength $g$.}
\label{fig:three}
\end{figure}

 One of the central results of this paper is that the formation of a BIC in the presence of SOC is physically robust phenomenon, because: First, one can again obtain a radiationless state by tuning the SOC strength while keeping the perturbed potential fixed, as shown in Fig.~\ref{fig:two}(c). The result is confirmed by Fig.~\ref{fig:three}, which shows the lifetime of dynamically excited states as a function of $\gamma$ obtained from the direct numerical solution of the equation $i\partial_t\bpsi=H\bpsi$ with the perturbed potential (\ref{potV}). The lifetime is defined as the time at which the peak amplitude of the wavepacket decreases $e$ times due to radiation leakage. In all cases, the exact BIC (\ref{BIC}) existing at $\gamma=0.5$ was used as input. In Fig.~\ref{fig:three}(a) the results shown with black dots were obtained for the potential (\ref{potV}) constructed at $\gamma=0.5$. Since the SOC strength is tunable (see~\cite{tun-SO} and references therein) we varied it while keeping the potential unchanged. As readily visible in the plot, a radiationless BIC exist only possible at $\gamma=0.5$, where the lifetime diverges. For all other values of the SOC strength the potential (\ref{potV}) does not support BICs, hence the wavepackets are always coupled to radiation modes and thus decays.  Second, similar phenomena are obtained for perturbed potentials $V_{\rm pert}$ with different perturbation amplitudes (red and green circles). Thus, for each potential there is a diverging peak that corresponds to a radiationless BIC, confirming the result shown in Fig.~\ref{fig:two}(c). Therefore, the important conclusion is that an exact shape of the potential is not required for the existence of BICs; rather, {\it there exists a broad family of both, exact and perturbed potentials that supports fully radiationless BIC states\/}.

\begin{figure}[t]
\includegraphics[width=1\columnwidth]{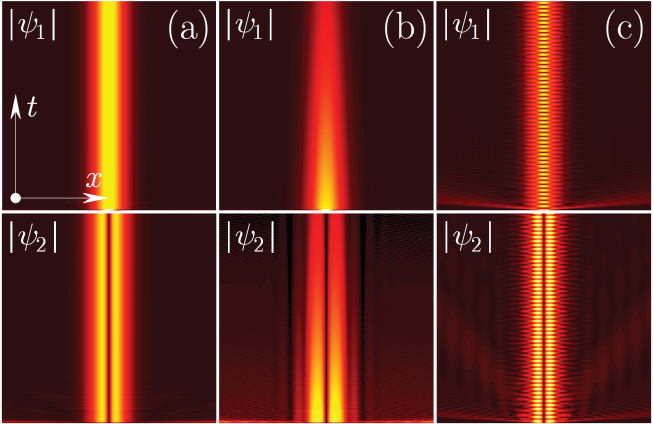}

\caption{(a) Excitation of a BIC in the potential designed to support it at $\gamma=0.5$, $\nu=0.7$. (b) Slow decay of a leaky state in the potential used in (a) when $\gamma=0.7$. (c) Simultaneous excitation of a coexisting BIC and standard guided modes in the potential (\ref{potV}) with $\gamma=1.3$, $\nu=0.7$, leading to beatings. The evolution is shown within the window $x\in [-20,20]$ for propagation up to $t=10^4$ in panels
	(a),(b) and up to $t =200$ in panel (c). }
\label{fig:four}
\end{figure}

Examples of evolution of the initially spin-polarized Gaussian wavepacket $\bpsi_1=\left(\exp(-x^2), 0\right)^T$ are depicted in Fig.~\ref{fig:four}. Panel (a) shows the excitation of a BIC in the potential (\ref{potV}) with $\gamma=0.5$. At such value of the SOC strength, the  potential supports only one mode - the BIC. Panel (b) shows the evolution of the same input but for $\gamma=0.7$ in the same potential (which was constructed to support a BIC for $\gamma=0.5$). No radiatioless modes exist in this case and one observes slow decay. Radiation occurs mostly from the second component. Panel (c) shows the case of the exact potential constructed for $\gamma=1.3$. Such potential supports not only the BIC, but also several modes from the discrete spectrum. Their simultaneous excitation results in irregular beatings.

\begin{figure}
\includegraphics[width=1\columnwidth]{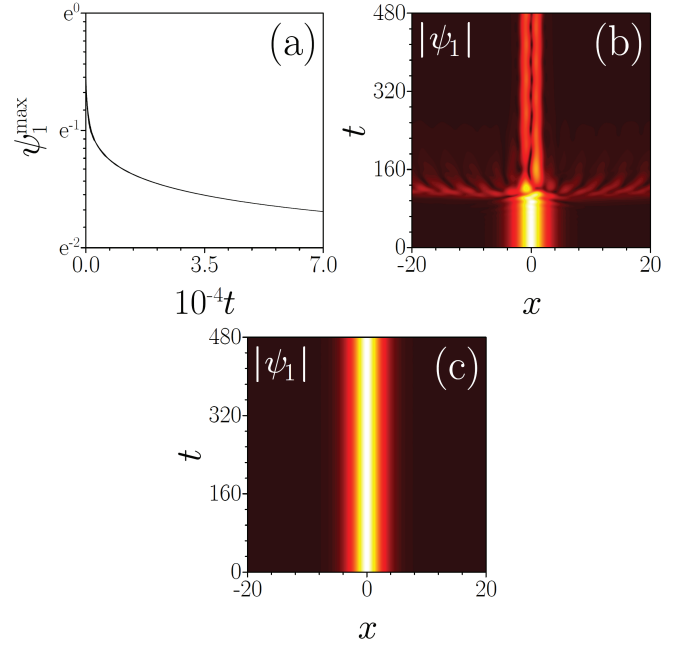}
\caption{  (a) Evolution of the peak amplitude of the $\psi_1$ component when the input is a Gaussian wavepacket in the potential (\ref{potV}) for $g=-2$, $\gamma=0.5$, $\nu=0.7$. Notice the logarithmic scale of the vertical axis. Evolution of the $\psi_1$ component of the perturbed nonlinear BIC in the potential (\ref{potVnl}) for $\gamma=1.4$ (b) and $\gamma=0.2$ (c) for $g=1$, $\nu=0.7$}
\label{fig:five}
\end{figure}

The previous cases correspond to linear systems, where BICs exist due to an interplay between SOC and the trapping potential. Next we elucidate whether BIC-like nonlinear modes may exist in BECs described by the one-dimensional spinor Gross-Pitaevskii equation
\begin{equation}
\label{GPE}
 i\bPsi_t=-\frac 12 \bPsi_{xx}-i\gamma \sigma_y\bPsi_x+\sigma_z \bPsi +\bV  \bPsi
 +g(\bPsi^\dagger\bPsi)\bPsi
\end{equation}
where $\bPsi$ is the dimensionless order parameters and $g$ describes two-body interactions (the sign of $g$ coincides with sign of the scattering length). First, we consider the impact of nonlinearity on the rigorous BICs (\ref{BIC}) that are exact solutions of the potential (\ref{potV}) when $g=0$. The numerical solution of (\ref{GPE}) using a standard Crank-Nicolson scheme shows that for a nonzero value of $g$ the evolution is always accompanied by radiation, thus leading to a finite lifetime. Figure~\ref{fig:three}(b) shows an example. The decay of peak amplitude is not exponential and it slows down when the amplitude decreases, as visible in Fig.~\ref{fig:five}(a).  However, using the algorithm described above for linear systems one can still construct nonlinear BICs. An example of a potential supporting such states is given by
\begin{eqnarray}
	\label{potVnl}
	\bV (x)=\frac{1}{\cosh^2(\nu x)}
	\left(\begin{array}{cc}
	U_1&0 \\ 0&U_2
	\end{array}\right)
	+\frac{g}{\cosh^4(\nu x)}
	\left(\begin{array}{cc}
	1&0 \\ 0&1
	\end{array}\right)
 \end{eqnarray}
where the parameters have to be chosen as
 \begin{eqnarray*}
	U_{1}=2\sqrt{1-\gamma^2\nu^2}-2-\nu^2-g-\frac{\gamma^2\nu^2 g}{ (\sqrt{1-\gamma^2\nu^2}-1)^2}
	 \\
  U_{2}=\frac{3\gamma^2\nu^4}{(\sqrt{1-\gamma^2\nu^2}-1)^2}+\frac{ 2(g+3\nu^2) }{ \sqrt{1-\gamma^2\nu^2}-1}.
\end{eqnarray*}
The respective BIC mode reads
\begin{eqnarray}
	\bPsi=\frac{\exp(-i\mu_\bic t)}{\cosh(\nu x)}\left(\begin{array}{c}
	\gamma\nu/(1-\sqrt{1-\gamma^2\nu^2})
	\\
	\tanh(\nu x)
	\end{array}\right)
\end{eqnarray}
where $\mu_\bic$ is given by Eq. (\ref{mu_bic}) with $\Omega=1$.
As expected on physical grounds, numerics show that such nonlinear BICs, supported by the potential (\ref{potVnl}), are unstable. Under the action of small perturbations they transform either into modes from the discrete spectrum [Fig. \ref{fig:five}(b)] or into dynamically oscillating patterns, depending on the value of $\gamma$. However, for a repulsive nonlinearity the strength of the instability decreases as the SOC strength decreases. Thus, for $\gamma=0.2$ no signs of instability are visible upon evolution up to $t=500$, as illustrated in Fig. \ref{fig:five}(c). The instability is suppressed, or at least drastically reduced also when $\gamma\to 1/\nu$ for both repulsive and attractive interactions. In both latter cases, the lifetime of the nonlinear BICs may be remarkably long --in practice indistinguishable from a rigorous radiationless state-- making possible their experimental excitation.

In conclusion, we have shown that the interplay between SOC, Zeeman splitting and the shape of the trapping potential may lead to the formation of BICs in atomic systems. While Zeeman splitting opens a semi-gap between two branches of the spectrum, SOC couples the modes and thereby becomes the key tuning parameter in the formation of BICs. Importantly, the phenomenon is robust, in the sense that a broad family of potentials exist where BICs may be excited by properly tuning the SOC strength. Several generalizations of this work are anticipated. First, the requirement of the exponential decay of the potential may be relaxed, so that, e.g., algebraically-decaying BICs may occur. Second, other physical types of SOC, such as Rashba and Dresselhaus, and even complex guage potentials, may be considered.  Third, extension of the analysis to other multilevel systems is possible. These include multicomponent BEC mixtures, SOC electron dynamics with applications for spintronics~\cite{spintronics}, exciton-polariton condensates with SOC~\cite{polariton}, among others.


\end{document}